\documentclass[conference]{IEEEtran}
\IEEEoverridecommandlockouts
\usepackage{cite}
\usepackage{amsmath,amssymb,amsfonts}
\usepackage{bbm}
\usepackage{algpseudocode} 
\usepackage{graphicx}
\usepackage{textcomp}
\usepackage{xcolor}
\usepackage{ifthen}
\usepackage{url}
\usepackage{tabularx}
\usepackage{algorithm} 
\usepackage[utf8]{inputenc} 
\usepackage[T1]{fontenc}    
\usepackage{url}            
\usepackage{amsfonts}       
\usepackage{nicefrac}       
\usepackage{microtype}   
\usepackage{amsmath}
\usepackage{subcaption}

\usepackage{amsmath}
\usepackage{booktabs}

\usepackage{mathrsfs}
\usepackage{tikz}
\usetikzlibrary{positioning}   
\usetikzlibrary{arrows.meta}   
\usetikzlibrary{fit}           
\usetikzlibrary{shapes.geometric, shapes.misc} 
\usepackage{subcaption}        

\def\BibTeX{{\rm B\kern-.05em{\sc i\kern-.025em b}\kern-.08em
    T\kern-.1667em\lower.7ex\hbox{E}\kern-.125emX}}

\usepackage[compact]{titlesec}
    \titlespacing{\section}{0pt}{1.0ex}{0.5ex}
    \titlespacing{\subsection}{0pt}{0.5ex}{0.5ex}
    \titlespacing{\subsubsection}{0pt}{1.0ex}{0.5ex}

\usepackage[letterpaper, left=0.6in, right=0.6in, bottom=1.095in, top=0.73in]{geometry}
\usepackage{framed}
\begin{document}

\title{Temporally Encoded Double DQN for Proactive PRB Allocation in O-RAN Enabled Industrial Networks
}
\author{Elahe Delavari, Xingqi Wu, and Junaid~Farooq\\
$^\dagger$Department of Electrical and Computer Engineering,
University of Michigan-Dearborn,\\ Dearborn, MI, 48128 USA, Emails: \{elahed, xingqiwu, mjfarooq\}@umich.edu.
}

\maketitle

\begin{abstract}
Fifth-generation (5G) wireless systems are increasingly adopted in smart manufacturing to support heterogeneous industrial workloads through services such as enhanced Mobile Broadband (eMBB) and Ultra-Reliable Low-Latency Communication (URLLC). However, industrial traffic is inherently process-driven and temporally correlated. So, static or reactive schedulers in the Open Radio Access Network (O-RAN) are inadequate for such non-stationary conditions, leading to sub-optimal utilization and violation of latency–reliability guarantees. This paper proposes a temporal-aware deep reinforcement learning (DRL) xApp for proactive Physical Resource Block (PRB) allocation in O-RAN-enabled industrial networks. The proposed framework integrates a long short-term memory (LSTM) encoder within a Double Deep Q-Network (DQN) to model sequential dependencies among slice-level Key Performance Indicators (KPIs), enabling predictive and stable decision-making. A continuous-time Markov chain (CTMC) traffic model is incorporated to emulate machine concurrency and process burstiness.
Experimental results show that the LSTM–Double DQN improves slice satisfaction, and buffer stability under moderate and heavy load, with the longest sequence window providing the strongest gains.

\end{abstract}


\section{Introduction}

\begin{figure}
\centering
\includegraphics[width=0.90\linewidth]{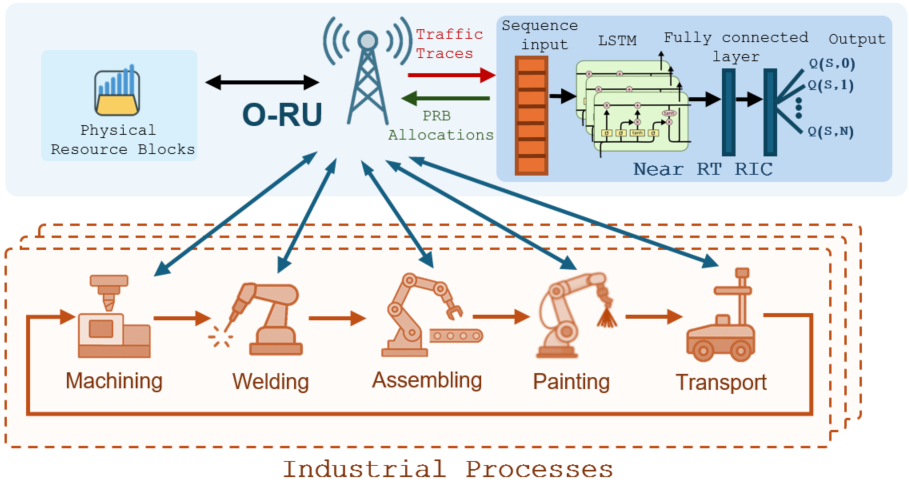}
\caption{O-RAN–enabled industrial network architecture with closed-loop PRB allocation by the Near-RT RIC.}
\label{fig:system_model}
\end{figure}

Smart manufacturing environments are increasingly composed of interconnected cyber-physical systems wherein machines, sensors, and robots coordinate over wireless links~\cite{noorarahim_smart_2022}. In such industrial Internet of Things (IIoT) settings, network traffic is highly heterogeneous, i.e., high-throughput services such as vision inspection and digital twin synchronization coexist with latency-sensitive control loops that demand ultra-reliable low-latency communication (URLLC)~\cite{ericsson_smart_2019}. These requirements evolve dynamically with production stages as shown in Fig.~\ref{fig:system_model}, rendering static or wired configurations inadequate for adapting to time-varying workloads~\cite{ludwig_reference_2022}. This motivates the use of learning-based schedulers, particularly reinforcement learning (RL), which can optimize decision-making over time horizons rather than rely on static allocation policies.


Fifth-generation (5G) networks support heterogeneous services through network slicing (NS), which enables multiple logical networks to coexist on shared infrastructure. Typical slice types include enhanced mobile broadband (eMBB), URLLC, and massive machine-type communication (mMTC)~\cite{wijethilaka_survey_2021}. Although this architecture facilitates service isolation and quality-of-service (QoS) differentiation, industrial workloads are often structured by machine-cycle–driven temporal dependencies. Fixed physical resource block (PRB) quotas and reactive slice scheduling are therefore insufficient, frequently resulting in degraded latency performance and inefficient spectrum use.


The Open Radio Access Network (O-RAN) architecture introduces openness and intelligence into the radio access infrastructure, enabling programmable control through the near-real-time RAN Intelligent Controller (Near-RT RIC)~\cite{RIC-1, RIC-2, RIC-3}. This architecture supports the deployment of artificial intelligence (AI)–driven applications (xApps) that can implement closed-loop control strategies tailored to dynamic and time-correlated traffic conditions~\cite{polese_understanding_2023}. In the context of smart factories, where workload fluctuations follow predictable production cycles, O-RAN provides a suitable platform for deploying temporally-aware control policies.

Conventional resource management strategies, such as proportional-fair scheduling and convex optimization, rely on instantaneous observations and are thus limited in settings where traffic patterns exhibit temporal correlation. Deep reinforcement learning (DRL) has emerged as a promising alternative for autonomous resource orchestration, with approaches such as Q-learning~\cite{shi2020reinforcement}, Deep Q Networks (DQN)~\cite{suh2022deep}, and actor–critic methods~\cite{alsenwi2021intelligent} outperforming traditional heuristics. However, most existing DRL-based xApps operate in a memoryless fashion, failing to exploit the short-term temporal regularities inherent in industrial traffic. To mitigate this limitation, recent studies have incorporated temporal modeling via long short-term memory (LSTM) encoders~\cite{li_lstm-based_2020} or predictive LSTM augmentations~\cite{li_lstm-characterized_2022,lotfi2023open,cai_deep_2024}. These methods improve performance in dynamic or partially observable scenarios but have not been extensively applied to proactive slice control in deterministic, process-driven industrial networks.

In this paper, we propose a temporally encoded DRL xApp for proactive PRB allocation in O-RAN–enabled smart manufacturing networks. The proposed method integrates an LSTM encoder within a Double DQN architecture to capture short-term temporal dependencies in slice-level KPIs such as PRB share, throughput, and buffer occupancy. The resulting temporal embedding enables predictive PRB allocation decisions that improve latency compliance and buffer stability under dynamic workloads.
We evaluate the proposed LSTM–Double DQN framework using an extended AI-RAN simulator incorporating a continuous-time Markov chain (CTMC) traffic model that emulates process-level concurrency. Experimental results demonstrate that temporal encoding improves slice satisfaction and significantly reduces queuing delay under moderate and heavy loads. The findings highlight the importance of temporal modeling in DRL-based O-RAN control applications for industrial settings.

\section{System Model} \label{sec:system_model}

We consider an O-RAN–enabled industrial network that serves heterogeneous wireless workloads generated by concurrently active machines on a factory floor. As shown in Fig.~\ref{fig:system_model}, the Near-RT RIC hosts the proposed temporal-aware deep DRL xApp that monitors slice-level traffic metrics and dynamically allocates PRBs to minimize end-to-end latency across services.

\subsection{Network Architecture and Slicing}

The system consists of a single O-RU providing connectivity to $M$ industrial user equipments (UEs), denoted $\mathcal{M}=\{1,2,\dots,M\}$. Each UE belongs to one of two service slices: eMBB or URLLC denoted as $\mathcal{M}_E $ and $\mathcal{M}_U$. The eMBB slice supports high-throughput flows such as video streaming or digital-twin synchronization, while the URLLC slice carries short control and actuation packets requiring minimal delay.
The total available bandwidth $B_{\mathrm{tot}}$ is divided into $K$ orthogonal PRBs, each of bandwidth $B$. At each time step $t\in\mathcal{T}=\{1,2,\dots,T\}$, the Near-RT RIC determines a PRB allocation matrix $\mathbf{E}(t)=[e_{m,k}(t)]$, where $e_{m,k}(t)\in\{0,1\}$ indicates whether PRB $k$ is assigned to UE $m$, where $\sum_{m\in\mathcal{M}} e_{m,k}(t)\leq1, \forall k\in\{1, 2, ..., K\}$. We then define $P_E(t)$ and $P_U(t)$ as the sum of the PRB allocation at time slot $t$ over the set of eMBB and URLLC users, respectively (i.e., $P_s(t)=\sum_{m \in \mathcal{M}_s}\sum_{k=1}^K e_{m,k}(t)$, where $s\in\{E,U\}$), and $P_E(t) + P_U(t) \le K$. The O-RU enforces this allocation within the subsequent transmission interval.

\subsection{Traffic Generation and Queueing Dynamics}

At time $t$, a subset $\mathcal{A}(t)\subseteq\mathcal{M}$ of machines is active, generating uplink or downlink traffic. Each active UE $m\in\mathcal{A}(t)$ produces packets of instantaneous size $\phi_m(t)$, which depends on its operational state and service type:
\begin{equation}
\phi_m(t) \sim
\begin{cases}
\mathcal{F}_E, & m\in\mathcal{M}_E,\\[2pt]
\mathcal{F}_U, & m\in\mathcal{M}_U,
\end{cases}
\end{equation}
where $\mathcal{F}_E$ and $\mathcal{F}_U$ denote the packet-length distributions of eMBB and URLLC traffic, respectively. Arriving packets are buffered at the O-RU in slice-specific queues $\mathcal{Q}_s(t)$ for $s\in\{E,U\}$.
The queues at the base station evolves as follows:
\begin{equation}
\mathcal{Q}_s(t{+}1)= \max\!\left[0,\, \mathcal{Q}_s(t) - \mathcal{T}_s \mathcal{R}_s(t)\right] + \!\!\sum_{m\in\mathcal{M}_s} \!\!\phi_m(t),
\label{eq:queue}
\end{equation}
where $\mathcal{T}_s$ is the scheduling interval and $\mathcal{R}_s(t)$ is the aggregate transmission rate achieved by slice $s$ given the PRBs assigned at time $t$.

\subsection{Data Rate Computation and PRB Demand}
\label{sec:prb_demand}
The achievable data rate for each UE is determined by its selected modulation and coding scheme (MCS) level, which depends on its Channel Quality Indicator (CQI). The contribution of PRB $k$ to UE $m$'s rate is given by:
\begin{equation}
\mathcal{R}_{m,k}(t)=
\frac{\mathcal{O} \times \mathcal{C} \times \mathcal{N}_{\mathrm{RE}}}{D_{\text{slot}}}\, e_{m,k}(t),
\end{equation}
where $\mathcal{O}$ is the modulation order and $\mathcal{C}$ is the effective coding rate associated with the selected MCS entry, $\mathcal{N}_{\text{RE}}$ is the number of resource elements per PRB per slot, and $\mathcal{D}_{\text{slot}}$ is the slot duration.
The total rate for UE $m$ satisfies
$
\mathcal{R}_m(t)=\sum_{k=1}^{K} \mathcal{R}_{m,k}(t).
$
The aggregate rate for slice $s$ is $\mathcal{R}_s(t)=\sum_{m\in\mathcal{M}_s}\mathcal{R}_m(t)$.
The slice throughput $m_s(t)$ denotes the total downlink data rate achieved by
slice $s$ at time $t$, computed as 
$m_s(t) = \sum_{m \in \mathcal{M}_s} \mathcal{R}_m(t)$.


The downlink PRB demand for each UE is derived from its MCS, Guaranteed Bit Rate (GBR) contract, and instantaneous buffer backlog. 
A UE with downlink GBR requirement $\mathcal{G}_m$ (in bps) is assigned a minimum PRB floor equal to
$
\left\lceil \frac{\mathcal{G}_m}{\mathcal{R}_{m,k}} \right\rceil.
$
In parallel, the O-RU monitors the UE’s downlink buffer, whose backlog $\mathcal{Q}_m$ (bytes) induces a rate demand 
$
I_m = \frac{8 \mathcal{Q}_m}{\mathcal{T}_s},
$
This translates into a backlog-driven PRB requirement equal to
$
\left\lceil 
\frac{\mathcal{I}_m}{\mathcal{R}_{m,k}}
\right\rceil.
$
To avoid allocating PRBs due to very small or transient bursts, we enforce the GBR floor only when the UE buffer holds at least one scheduler-step’s worth of GBR traffic. In practice, this means applying the GBR constraint when
$
\mathcal{Q}_m \ge \frac{\mathcal{G}_m \mathcal{T}_s}{8} \quad \text{and} \quad \left\lceil 
\frac{\mathcal{I}_m}{\mathcal{R}_{m,k}}
\right\rceil > 0.
$
Under this condition, the demand is raised to the minimum number of PRBs needed to satisfy the GBR; otherwise, no GBR floor is applied. The final PRB requirement reported to the xApp is therefore
$
D_m =
\max\left(
\left\lceil \frac{\mathcal{I}_m}{\mathcal{R}_{m,k}} \right\rceil,\;
\left\lceil \frac{G_m}{\mathcal{R}_{m,k}} \right\rceil
\right)
$
which ensures that sustained GBR obligations are met while remaining responsive to real-time queue dynamics.

\subsection{Latency Evaluation and Performance Characterization}

For each slice $s\in\{E,U\}$, the instantaneous latency at time $t$ is derived
from the relationship between the slice backlog and the serving rate of the
gNB. The resulting slice
latency is
\begin{equation}
    L_s(t)=\frac{\mathcal{Q}_s(t)}{\mathcal{R}_s(t)}.
\end{equation}

To characterize how effectively a slice uses its PRB allocation to maintain
acceptable delay, we introduce a slice-level \emph{PRB-efficiency} measure. In
our model, efficiency depends not only on the PRB budget but also on how well
the slice meets its latency target. A general expression is
\begin{equation}
\label{eq:reward}
    \eta_s(t)
    =
    f\!\left(
        \frac{\psi_s\!\big(L_s(t)\big)}{ P_s(t)}
    \right),
\end{equation}
where $P_s(t)$ is the PRB quota assigned to slice $s$, and 
$\psi_s : \mathbb{R} \rightarrow [0,1]$ is a latency-alignment function that 
assigns higher values when the observed slice latency $L_s(t)$ approaches 
the slice-specific latency target $L_s^\star$ (much stricter for URLLC than 
for eMBB). 
The monotone function $f(\cdot)$ models diminishing returns with respect to PRB
consumption. Consequently, $\eta_s(t)$ reflects how efficiently slice~$s$
maintains its latency objective given the PRB resources allocated to it.
\subsection{Optimization Objective}
The Near-RT RIC
adjusts $\{P_E(t), P_U(t)\}$ at every scheduling interval according to the
xApp's control policy. And both slices are evaluated through a latency-alignment measure that is
normalized by their PRB usage, the PRB-control problem aims to allocate
sufficient resources to keep slice latency near its target while avoiding unnecessary spectrum consumption. This trade-off is captured through the
efficiency indicators $\eta_s(t)$ introduced above, which increase when latency
is well aligned and decrease when excessive PRBs are used.
The high-level PRB allocation problem is therefore described as follows:
\begin{equation}
    \max_{\{P_s(t)\}}
    \; G\!\left(
        \eta_E(t),\, \eta_U(t)
    \right),
\end{equation}
where $G(\cdot)$ is a non-decreasing aggregation function representing overall system performance. Direct optimization of $G$ is generally intractable due to the stochastic and queue-coupled dynamics. Therefore, a surrogate reward formulation is introduced in Section~\ref{sec:methodology} to enable tractable reinforcement learning while preserving the underlying trade-offs.


\section{Methodology} \label{sec:methodology}

To enable predictive and latency-aware resource allocation in O-RAN–enabled industrial networks, we propose a RL–based control framework that dynamically adjusts PRB allocations across network slices. This section describes the RL formulation, state and action design, reward function, temporal encoding architecture, and training procedure of the proposed LSTM–Double DQN xApp.
\subsection{Reinforcement Learning Formulation}
The PRB allocation task is modeled as a sequential decision-making problem, where each action influences future slice latencies and buffer states. Due to short-term temporal dependencies in industrial traffic, the learning agent must exploit past state information to perform stable control under dynamic workloads. To this end, we adopt a Double Deep Q-Network (Double DQN) framework augmented with a Long Short-Term Memory (LSTM) encoder. This temporally aware design enables the xApp to capture evolving traffic patterns without requiring explicit prediction.

We additionally implement a memoryless baseline using a multilayer perceptron (MLP)–Double DQN, where the instantaneous six-dimensional network state is processed through two fully connected layers. This baseline is used to quantify the performance gain achieved by temporal modeling in the LSTM–Double DQN agent.
The problem is formalized as a Markov Decision Process (MDP) defined by the tuple $(\mathcal{S}, \mathcal{A}, \mathcal{P}, \mathscr{R}, \gamma)$, where $\mathcal{S}$ denotes the slice-level network state, $\mathcal{A}$ is the action space, $\mathcal{P}$ represents the environment dynamics, $\mathscr{R}$ is the reward function, and $\gamma$ is the discount factor. The xApp observes the current state, selects an action to adjust PRB allocations, and receives a reward based on slice-level QoS metrics. Based on this MDP formulation, the components of the state space, action space, and reward function are defined as follows.

\noindent \textbf{1) State Space:}
At time $t$, the agent observes
$
s_t = [\tilde{P}_E, \tilde{m}_E, \tilde{\mathcal{Q}}_E, \tilde{P}_U, \tilde{m}_U, \tilde{\mathcal{Q}}_U],
$
where $\tilde{P}_s$ is the normalized PRB share, $\tilde{m}_s$ the normalized throughput, and $\tilde{\mathcal{Q}}_s$ the normalized buffer backlog for slice 
$s\in\{E, U\}$. All features lie in $[0,1]$ for numerical stability.

\noindent \textbf{2) Action Space:}
Each action adjusts slice PRBs in small directional increments
$
a_t = (\Delta p_E, \Delta p_U), \quad \Delta p_s \in \{-1, 0, +1\}.
$
The effective change is $\Delta P_s = \lambda \Delta p_s$, where $\lambda$ is the step size. The resulting actions increase, decrease, or maintain PRBs for each slice
; any over-allocation is automatically rescaled to satisfy the PRB budget.

\noindent \textbf {3) Reward Function:}
The reward promotes low-latency operation and efficient PRB usage. 
For each slice $s\in\{E, U\}$ slice latency $L_s$ relative to its target $L_s^\star$ is mapped through a logistic scoring function,
\begin{equation}
\psi_s = \frac{1}{1 + \exp\!\left[-a_s (L_s - L_s^\star) - b_s\right]} ,
\end{equation}

where $a_s$ and $b_s$ control the slope and offset of the latency penalty. 
Efficiency is incorporated as:
\begin{equation}
\mathscr{R}_s = f\!\left( \frac{\alpha_s\ . \psi_s(L_s)}{\sigma_s + P_s} \right),
\end{equation}

where $P_s$ is the current PRB quota. 
Here, $\sigma_s > 0$ is a small constant that prevents the denominator from vanishing and 
$\alpha_s$ balances the latency score with the PRB  term.
The global reward is a weighted sum,
$
\mathscr{R} = w_E \mathscr{R}_{\text{E}} +  w_U \mathscr{R}_{\text{U}}
$
This formulation emphasizes URLLC reliability while encouraging balanced and proactive PRB allocation.
Although $G(\eta_E,\eta_U)$ is not optimized directly, the reward is 
constructed as a tractable surrogate of the same objective. Each slice-level 
term $\mathscr{R}_s$ reflects the efficiency structure of $\eta_s$ through latency 
alignment and PRB normalization, and the combined reward 
$\mathscr{R}$ serves as an RL-friendly approximation of $G(\cdot)$.


\subsection{Double DQN Training Procedure}
The PRB allocator is trained using the Double DQN, which reduces the overestimation bias of standard DQN by separating action selection and action evaluation. Two networks are maintained: an online network $Q(s,a;\theta)$ and a target network $Q(s,a;\theta^{-})$ periodically synced with the online parameters. At each update, a transition $(s,a,r,s',d)$ is sampled from the replay buffer, where $d \in \{0,1\}$ is the terminal flag indicating whether $s'$ corresponds to the end of an episode.
The next action is chosen using the online network and evaluated using the target network. The Double DQN target becomes
\begin{equation}
\label{eq:ddqn_target}
y=\mathscr{R} + (1-d)\gamma\, Q(s',\arg\max_{a} Q(s',a;\theta);\theta^{-}).
\end{equation}



The online network parameters $\theta$ are updated by minimizing a Huber
loss on the temporal-difference (TD) error $\delta = y - Q(s,a;\theta)$,
using minibatch updates optimized with Adam optimizer. Exploration follows a
linearly decaying $\epsilon$-greedy policy, and transitions are sampled
from a replay buffer for stable training. 

The MLP–Double DQN processes the instantaneous six-dimensional
state through two fully connected layers of width 256 with ReLU activations.
A final linear layer outputs Q-values for the nine discrete PRB-adjustment
actions. This feedforward design corresponds to the memoryless agent used as
a baseline throughout our evaluation.

\subsection{LSTM-Double DQN for Temporally-Aware PRB Allocation}

\begin{figure}[t]
    \centering
    \includegraphics[width=0.98\linewidth]{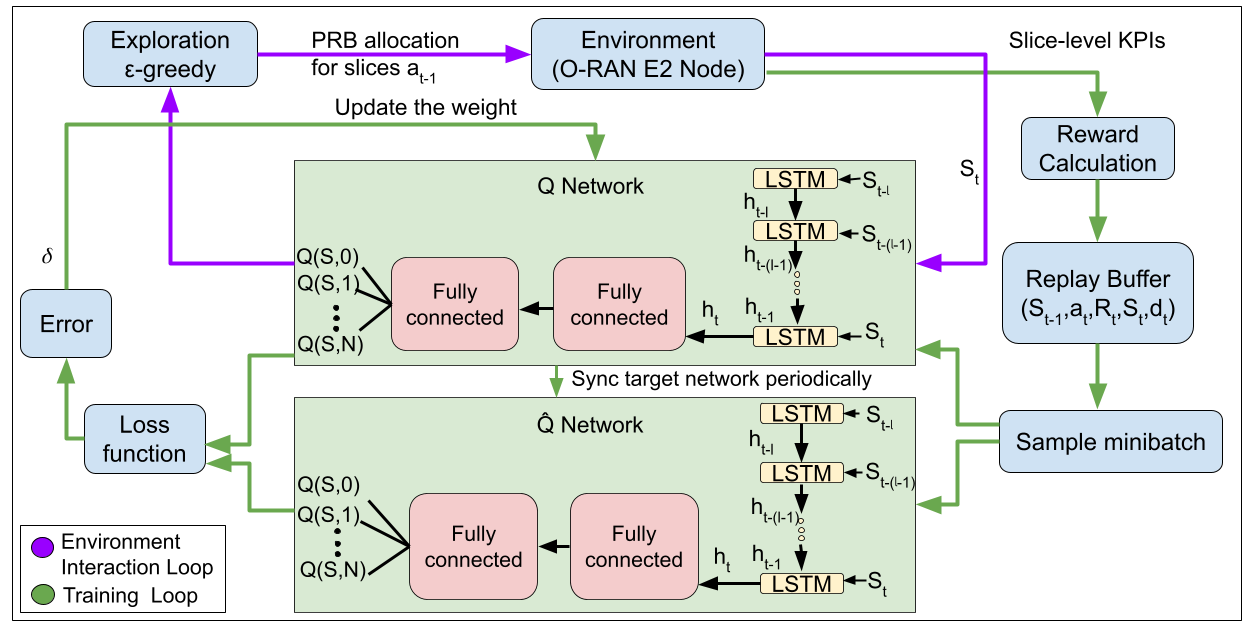}
    \caption{Proposed LSTM--Double DQN xApp.}
    \label{fig:lstm_dqn_arch}
\end{figure}

To exploit short-term temporal patterns in traffic, we extend the feedforward agent with an LSTM encoder. 
Instead of using only the instantaneous
six-dimensional state, the agent forms a sequence window
$S_t = (s_{t-\ell+1}, \ldots, s_t)$, where $\ell$ is the sequence
length hyperparameter that specifies how many past slice-level states are
included in the window. This sequence is passed to a single-layer LSTM:
$
h_t=\textit{LSTM}(S_t).
$
The vector $h_t$ serves as a
compact summary of the recent traffic evolution and is forwarded to a fully connected head that outputs Q-values for the nine PRB adjustment actions.
Only the final hidden state $h_t$ is used because it represents the
LSTM’s learned encoding of the entire sequence and acts as a
fixed-length temporal embedding suitable for Q-learning. 
To clarify the overall processing pipeline of the temporal encoder and the Double DQN, Fig.~\ref{fig:lstm_dqn_arch} illustrates how a window of the most recent $l$ slice-level states is passed through an LSTM to produce the temporal feature $h_t$, which is then fed into the fully connected Q-network.
Both online and target networks share the same LSTM-based encoder; the Double DQN update, target construction, and Huber-loss minimization remain identical to the feedforward case. The complete training loop is shown in Algorithm~\ref{alg:lstm_dqn}. 
The LSTM–Double DQN replaces the feedforward encoder with a single-layer
LSTM of hidden size 128 that ingests the sequence window $S_t$. All the other layers are the same as the MLP version.

\begin{algorithm}[t]
\caption{Episodic LSTM--Double DQN for PRB Allocation}
\label{alg:lstm_dqn}
\begin{algorithmic}[1]
\Require sequence length $\ell$, replay buffer $\mathcal{B}$, networks $Q(\cdot;\theta)$ and $Q(\cdot;\theta^{-})$, 
discount $\gamma$
\State Initialize $\theta, \theta^{-}$
\For{each episode $e$}
    \State Reset env, observe $s_0$, set window $\mathcal{W}\!\leftarrow\!\{s_0\}$
    \For{each step $t$}
        \State Form sequence $S_t$ from last $\ell$ states in $\mathcal{W}$
        \State Select $a_t$ via $\epsilon$-greedy on $Q(S_t,a;\theta)$
        \State Execute $a_t$, observe $s_{t+1}$, $\mathscr{R}_t$, $d_t$
        \State Append $s_{t+1}$ to $\mathcal{W}$ (keep last $\ell$)
        \State Form $S_{t+1}$ from updated $\mathcal{W}$
        \State Store $(S_t,a_t,\mathscr{R}_t,S_{t+1},d_t)$ in $\mathcal{B}$
        \If{ready to train}
            \State Sample minibatch $(S,a,\mathscr{R},S',d)$ from $\mathcal{B}$
            \State Compute $a'=\arg\max_a Q(S',a;\theta)$
            \State Compute target $y=\mathscr{R}+(1-d)\gamma\,Q(S',a';\theta^{-})$
            \State Update $\theta$ by minimizing $\mathrm{Huber}(Q(S,a;\theta),y)$
            \If{target update}
                $\theta^{-}\leftarrow\theta$\EndIf
        \EndIf
        \If{$d_t=1$} \State \textbf{break} \EndIf
    \EndFor
\EndFor
\end{algorithmic}
\end{algorithm}

\section{Performance Evaluation} \label{sec:results}

We build on the Python-based AI-RAN Simulator~\cite{airansim}, which emulates a 5G-compliant O-RAN system with gNBs, UEs, and a RIC. Our extended version\footnote{\url{https://github.com/ElaheDlv/Proactive-PRB-Allocation}} simulates a $2\,\text{km}\!\times\!2\,\text{km}$ region with a single gNB deployed at the center, serving one n78 cell using standard NR numerology. UEs are placed uniformly, and initial slice-level PRB allocations reflect training and testing configurations. Transport block sizes follow 3GPP TS~38.214, Table~5.1.3.1--2, using modulation orders $\mathcal{O} \in \{2,4,6,8\}$ and corresponding coding rates $\mathcal{C}$. Simulator parameters are summarized in Table~\ref{tab:sim_params}.


\subsection{Simulation Setup} \label{subsec:sim_setup}
Traffic is generated using an event-driven CTMC
that models the manufacturing processes executed by two industrial machines
connected to the O-RAN base station. 
Machine~1 performs a machining operation (M1), Machine~2 performs an assembly
operation (M2), and both may operate concurrently depending on the production
workflow. The CTMC state space is 
$\mathcal{S} = \{S_1, S_{12}, S_2\}$, 
where $S_1$ corresponds to machining (M1), $S_2$ corresponds to assembly (M2),
and $S_{12}$ represents the concurrent machining–assembly phase (M1+M2).
State transitions follow the rate matrix
\[
\mathscr{Q} = 
\begin{bmatrix}
-0.75 & 0.40 & 0.35\\
0 & -0.20 & 0.20\\
0.20 & 0 & -0.20
\end{bmatrix},
\]
where the exit rate from state $i$ is $\lambda_i=\sum_{j\neq i} q_{ij}$ and dwell times are sampled from $ \mathrm{Exp}(1/\lambda_i)$. Normalizing the rows of $\mathscr{Q}$ yields the transition probability matrix $\mathscr{P}$, ensuring statistically correct and memoryless state evolution.


As the CTMC evolves, each active interval is mapped to a traffic slice: machining generates URLLC traffic, while assembly generates eMBB traffic. URLLC arrivals follow an Erlang distribution $\mathrm{Erlang}(k{=}3,\ \mathrm{scale}{=}0.5\ \mathrm{ms})$ with a packet size distribution $\mathcal{F}_U$ as $\mathcal{N}(60,5)$. The eMBB traffic is generated using Lognormal inter-arrival times $\mathrm{Lognormal}(\mu=\ln 7,\ \sigma=0.55)$ and packet size distribution $\mathcal{F}_E$ is  $\mathrm{Lognormal}(6.0,0.5)$, producing heavy-tailed bursts. Packets are time-stamped according to the sampled inter-arrival intervals and queued at the base station buffer.
To generate diverse traffic traces, we vary two scaling parameters: $\alpha$ amplifies forward transitions ($S_1\!\rightarrow\! S_{12},S_2$) and $\beta$ scales return transitions ($S_2\!\rightarrow\! S_1$), applied to the corresponding entries of $Q$. Varying these parameters produces multiple trace sets with different concurrency levels. The resulting packet-level traces are aligned to a global timeline and imported into the O-RAN simulator for evaluating PRB-allocation policies.

\begin{table}[t]
\centering
\caption{Simulator Parameters and Reward Hyperparameters}
\label{tab:sim_params}
\begin{tabular}{c}
\begin{minipage}{0.47\columnwidth}
\centering
\textbf{Simulator Parameters}\\[2pt]
\begin{tabular}{lc}
\toprule
\textbf{Param.} & \textbf{Value} \\
\midrule
$B$ & $100$\,MHz \\
$f_c$ & $3.5$\,GHz \\
Cell radius & $800$\,m \\
Tx power & $40$\,dBm \\
$P_{\text{DL}}^{\max},\,P_{\text{UL}}^{\max}$ & $218,\;55$ \\
$\mathcal{N}_{\text{RE}}$ & $168$ \\
$\mathcal{D}_{\text{slot}}$ & $\approx 1$\,ms \\
$\mathcal{G}_E,\mathcal{G}_U$ & $10,1$\,Mbps \\
\bottomrule
\end{tabular}
\end{minipage}
\hfill
\begin{minipage}{0.47\columnwidth}
\centering
\textbf{Reward Parameters}\\[2pt]
\begin{tabular}{lc}
\toprule
\textbf{Param.} & \textbf{eMBB , URLLC} \\
\midrule
$L_s^\star$ & $0.1 , 0.001$\,s \\
$a_s$ & $-35 , -9000$ \\
$b_s$ & $4.5 , 4.5$ \\
$\alpha_s$ & $100 , 50$ \\
$\sigma_s$ & $5 , 2$ \\
$K$ & $5 , 5$ \\
$w_s$ & $0.5 , 0.5$ \\
\bottomrule
\end{tabular}
\end{minipage}
\end{tabular}
\end{table}



For reproducibility, the specific numerical
values used in our experiments are summarized in
Table~\ref{tab:sim_params}.
For the latency-alignment component, the function $\psi_s(L_s)$ is
instantiated using the logistic form in equation \ref{eq:reward},
which assigns higher scores when the slice latency approaches its target
$L_s^\star$. The shaping function appearing in the PRB-efficiency term is
realized as $f(x) = K \ln(1 + x)$,
where $K$ serves as a global scaling factor controlling the magnitude of the
reward. 

\begin{figure}[t]
    \centering
    \includegraphics[width=0.8\columnwidth]{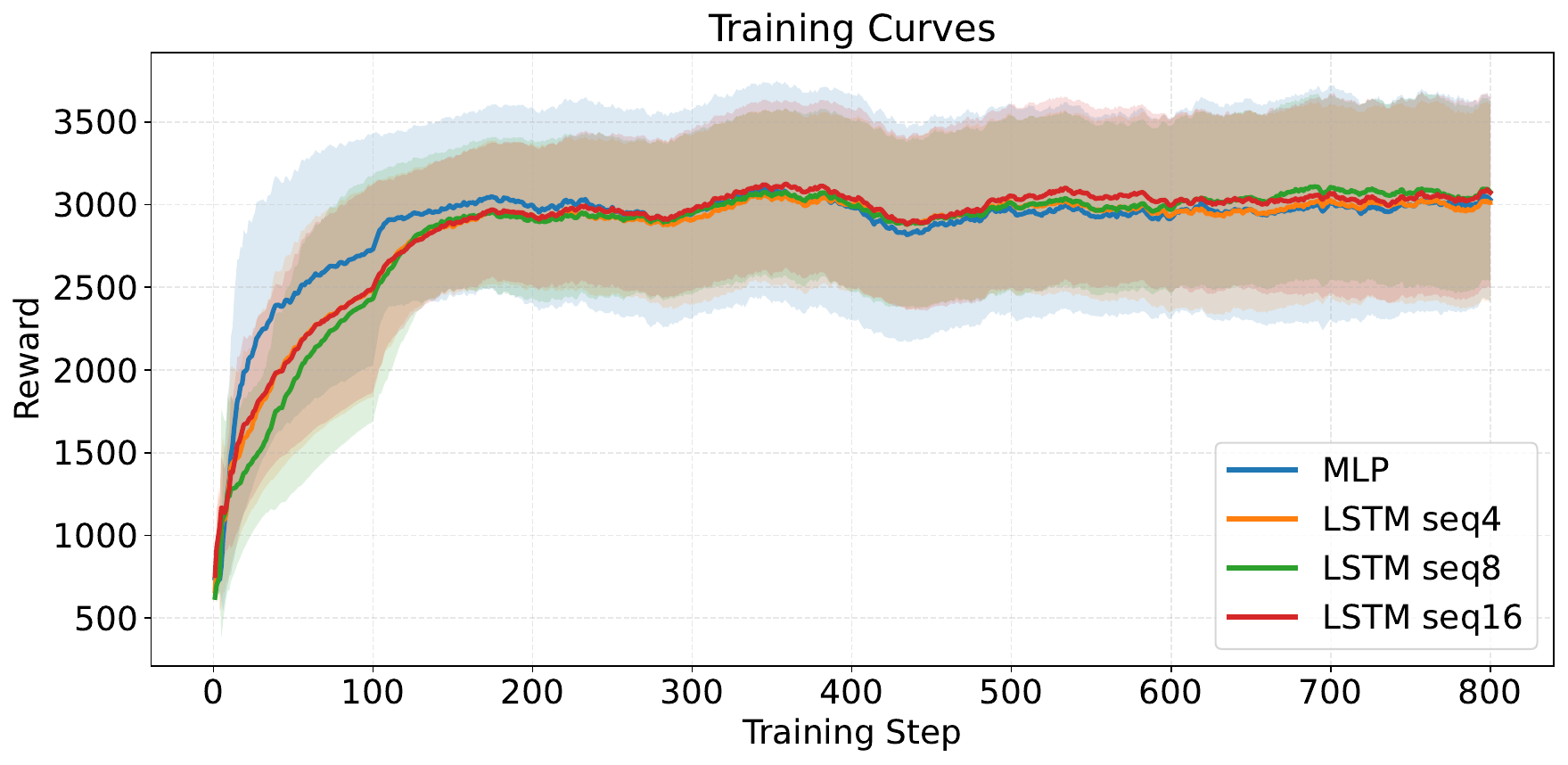}
    \caption{Training episodic return comparison for the MLP-Double DQN and LSTM-Double DQN agents.}
    \label{fig:training_return}
\end{figure}

\subsection{Experimental Setup}

Training was conducted over 800 episodes using randomized CTMC traces and variable load levels. Each episode spans 20 seconds of simulation time.

\noindent{1) \emph{Training Configurations}:}
The agent is exposed to diverse conditions by varying the number of active UEs
$\{1,3,5,7,9\}$, all combinations of initial PRB quotas for eMBB and URLLC slices
$\{5,7,9,11\}^2$, and CTMC scaling factors
$\alpha,\beta \in \{0.25,0.5,1,1.5,2\}$. For training, we use a discount factor of $\gamma = 0.99$, a learning rate of $5\times 10^{-4}$, and a PRB adjustment step size of $\lambda = 5$.

\noindent{2) \emph{Generalization Tests:}}
Generalization is evaluated on UE densities
$\{1,2,4,6,8\}$ using a fixed initial allocation of
$\text{eMBB\_PRBs}=\text{URLLC\_PRBs}=5$.  
These intermediate load levels test robustness without departing from the
training regime.


\subsection{Evaluation Metrics}

To assess the performance of the proposed scheduler under mixed eMBB and URLLC traffic, several quantitative metrics are monitored during simulation.  
These metrics capture both instantaneous and long term characteristics of satisfaction, buffer, and resource utilization at the base station.  
The following indicators are computed at each decision interval and aggregated over time to evaluate steady state performance:

\begin{enumerate}

\item {\emph{Slice Satisfaction Ratio:}}  
The satisfaction ratio $P_s/D_s$ quantifies how effectively each slice’s PRB demand is met, where $P_s$ and $D_s$ denote the granted and demanded PRBs, respectively.  
High satisfaction values indicate that the scheduler is able to allocate sufficient resources to maintain service quality under varying loads.

\item {\emph{PRB Allocation Efficiency:}}
It reflects how well each slice converts its allocated PRBs into
low-latency service. We compute efficiency using the same latency-based
scoring function and PRB-normalized shaping terms defined in the reward
formulation. This metric therefore captures both latency performance and the diminishing returns of allocating excessive PRBs to a slice.

\item {\emph{Average Buffer Size:}}
The average buffer size measures the aggregate downlink queue occupancy for the corresponding slice.
Large buffer values indicate congestion and insufficient PRB allocation,
whereas small values reflect smooth packet delivery with minimal queuing.
\end{enumerate}

\subsection{Simulation Results}
\begin{figure}[t]
    \centering
    \includegraphics[width=0.6\linewidth]{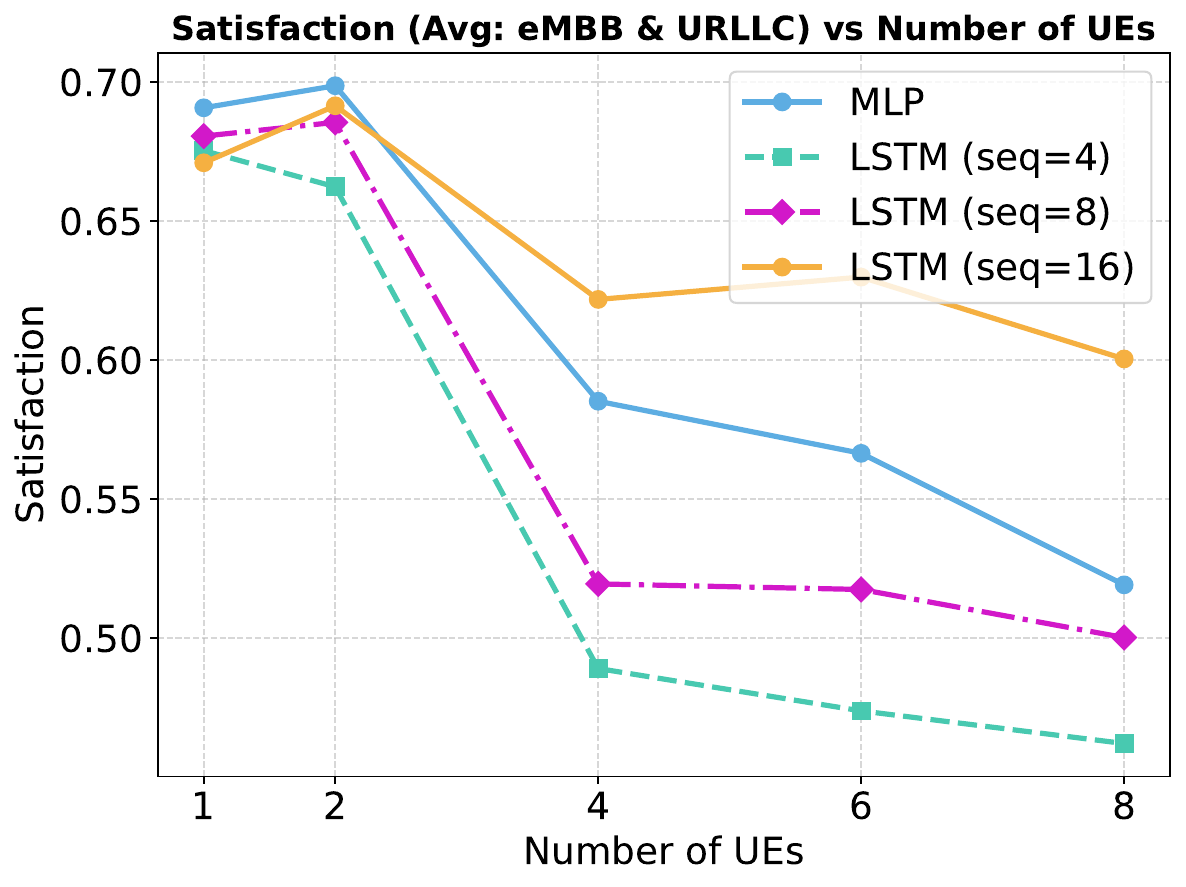}
    \caption{Satisfaction vs.\ number of UEs.}
    \label{fig:ue_satisfaction}
\end{figure}

\begin{figure}[t]
    \centering
    \includegraphics[width=0.6\linewidth]{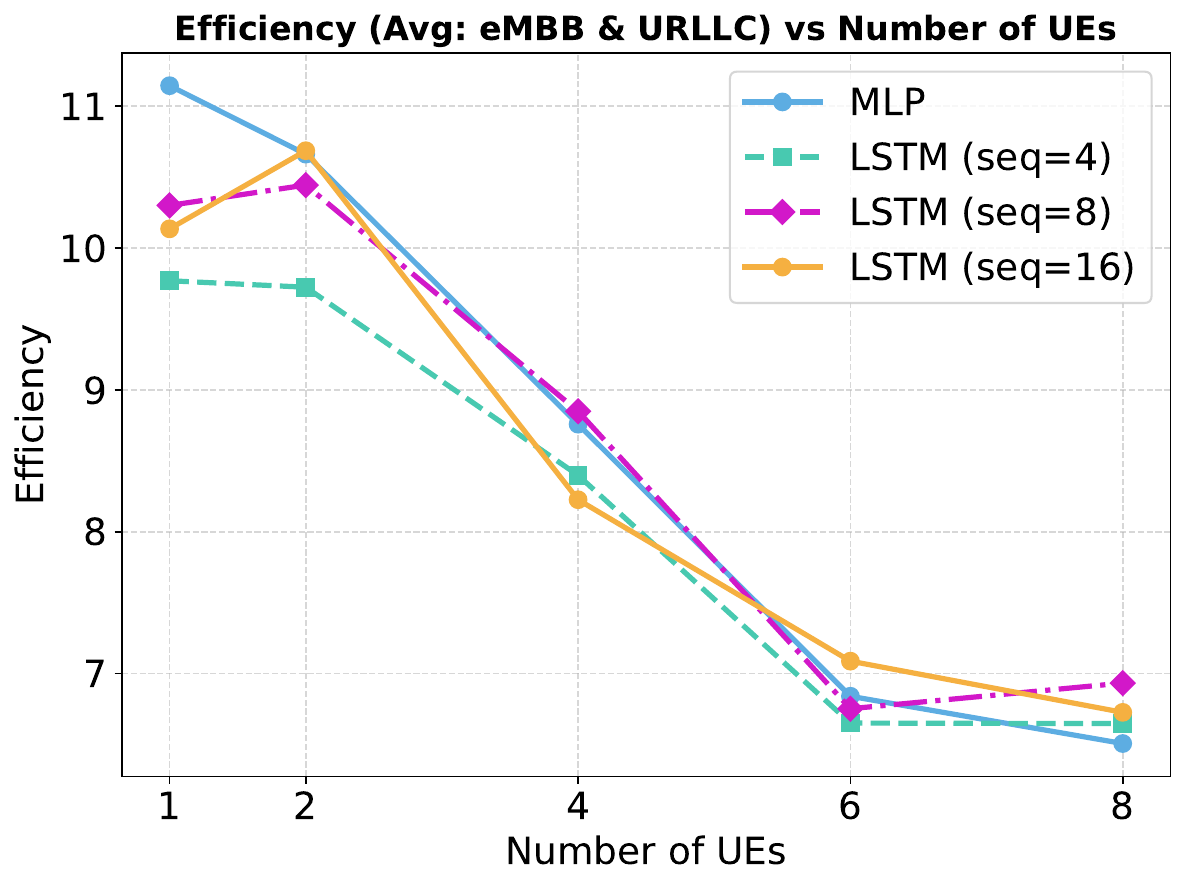}
    \caption{Efficiency vs.\ number of UEs.}
    \label{fig:ue_Efficiency}
\end{figure}

\begin{figure}[t]
    \centering
    \includegraphics[width=0.6\linewidth]{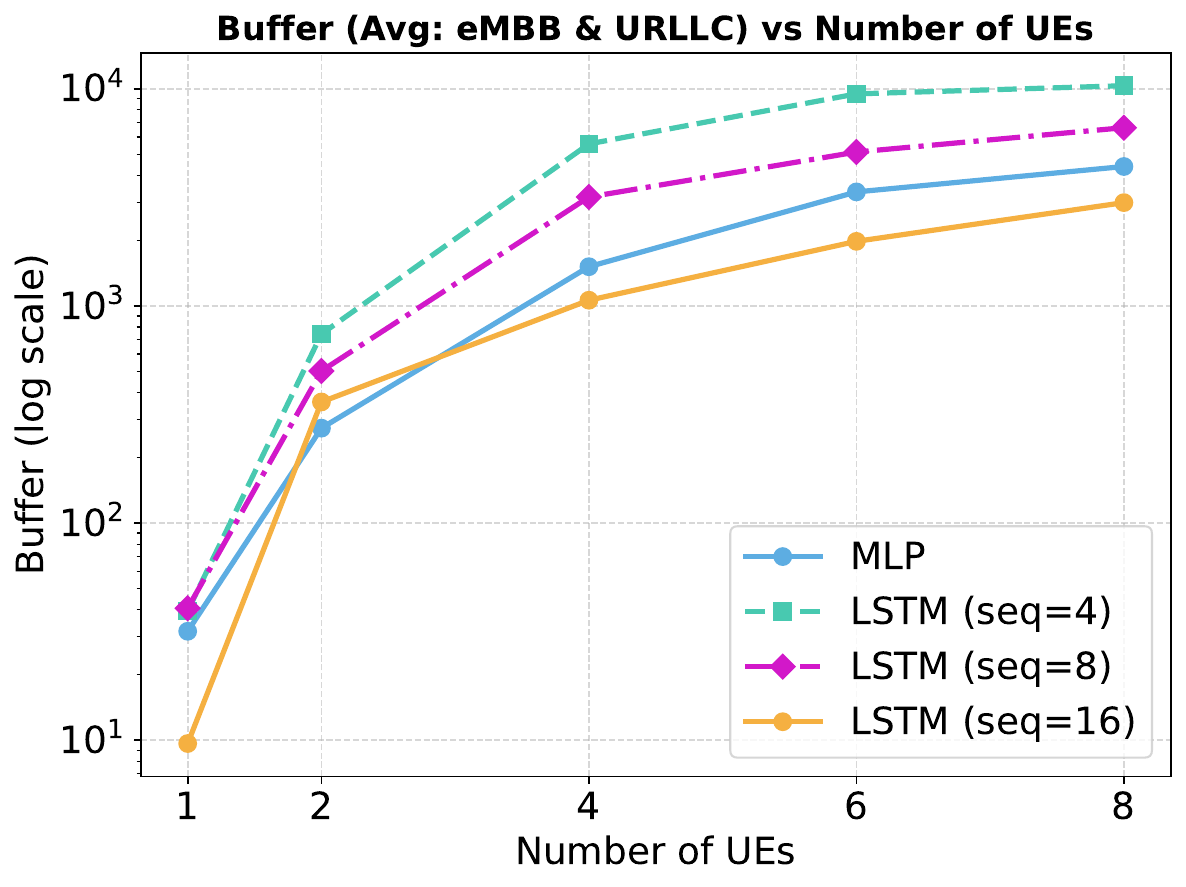}
    \caption{Buffer vs.\ number of UEs.}
    \label{fig:ue_Buffer}
\end{figure}

The performance of the proposed temporally encoded LSTM Double DQN xApp is evaluated using the test traces described earlier, with particular emphasis on the model’s behavior under previously unseen load levels and traffic temporal patterns. The results provide a detailed view of how temporal awareness improves resource allocation in the presence of bursty and non stationary industrial traffic.
Fig.~\ref{fig:training_return} compares the episodic return during training for the MLP–Double DQN and the LSTM variants with sequence lengths $\ell \in \{4, 8, 16\}$. All models converge stably, but the LSTM agents achieve higher returns, especially with longer sequences. The generalization results under unseen UE densities are presented in Figs.~\ref{fig:ue_satisfaction}–\ref{fig:ue_Buffer}. Satisfaction trends (Fig.~\ref{fig:ue_satisfaction}) show that all agents perform well under light load, but the MLP baseline deteriorates under congestion. The LSTM agent with $\ell{=}16$ achieves the highest satisfaction at moderate and heavy load, highlighting the benefits of long-term temporal context. 
PRB efficiency (Fig.~\ref{fig:ue_Efficiency}) remains comparable across agents at low to moderate load. However, under increased traffic intensity, the LSTM agents show improved efficiency consistency, suggesting that temporal awareness helps preserve latency without over-allocating PRBs. Buffer size (Fig.~\ref{fig:ue_Buffer}) reveals that MLP agents exhibit queue buildup under heavy load, whereas the LSTM variant with $\ell{=}16$ maintains significantly smaller buffers, indicating more proactive and anticipatory scheduling.

The results confirm that temporally encoded DRL agents provide improved stability and robustness compared to memoryless counterparts. In particular, the LSTM–Double DQN with longer sequence lengths ($\ell = 16$) consistently outperforms both the MLP and shorter LSTM variants across all metrics. Short sequences (e.g., $\ell = 4$) may fail to capture the CTMC structure, leading to noisy embeddings and suboptimal decisions. Overall, temporal encoding improves slice satisfaction, reduces congestion, and ensures balanced resource utilization in non-stationary industrial networks.

\section{Conclusion} \label{sec:conclusion}

This paper presented a temporally aware RL xApp for proactive PRB allocation in O-RAN–enabled industrial networks. By integrating an LSTM encoder into a Double DQN framework, the proposed scheduler captures short-term temporal patterns in slice-level KPIs and anticipates CTMC-driven traffic variations.
Experimental results show that the LSTM–Double DQN can improve slice satisfaction and reduces buffer buildup compared to the feedforward MLP baseline, particularly under moderate and heavy load. Longer sequence windows provide the strongest gains, while short windows capture insufficient temporal structure. These findings highlight the importance of temporal modeling for stable, resource-efficient PRB allocation in non-stationary industrial environments.
Future work includes extending the design to multiple slices, exploring adaptive sequence-length selection, and deploying the approach on near-real-time RIC platforms for large-scale validation.

\bibliographystyle{ieeetr} 
\bibliography{PRB_Allocation_LSTM} 
\end{document}